\documentclass[onecolumn,11pt,a4paper,aps]{revtex4}

\newcommand{\ch}{\mathrm{ch}}
\newcommand{\sh}{\mathrm{sh}}

\usepackage{amsmath}
\usepackage{graphics}
\usepackage{graphicx}
\usepackage{psfrag}

\begin{document}

\title{Entanglement in composite free-fermion systems}
\author{Viktor Eisler$^1$, Ming-Chiang Chung$^2$ and Ingo Peschel$^3$}
\affiliation{
$^1$MTA-ELTE Theoretical Physics Research Group, E\"otv\"os Lor\'and University,
P\'azm\'any P\'eter s\'et\'any 1/a, H-1117 Budapest, Hungary\\
$^2$Department of Physics, National Chung Hsing University, Taichung 40227, Taiwan\\ 
$^3$Fachbereich Physik, Freie Universit\"at Berlin,
Arnimallee 14, D-14195 Berlin, Germany
}

\begin{abstract}
We consider fermionic chains where the two halves are either metals with different
bandwidths or a metal and an insulator. Both are coupled together by a special bond. 
We study the ground-state entanglement entropy between the two pieces, its dependence 
on the parameters and its asymptotic form. We also discuss the features of the entanglement 
Hamiltonians in both subsystems and the evolution of the entanglement entropy after 
joining the two parts of the system.

\end{abstract}
\maketitle

\section{Introduction}

 The entanglement properties of free-fermion systems have been the topic of many studies
and various different cases have been investigated, see e.g. \cite{review09}. In one
dimension, these comprise homogeneous critical chains where the ground-state entanglement 
entropy $S$ varies logarithmically with the size $L$ of the subsystem, and non-critical 
ones where it approaches a constant for large $L$. Single defects, both at 
interfaces \cite{Peschel05,Igloi/Szatmari/Lin09,Eisler/Peschel10,Eisler/Garmon10,
CMV11,CMV12a,CMV12b,Peschel/Eisler12,Vasseur/Jacobsen/Saleur14} 
and in the interior 
\cite{Eisler/Peschel07,Ossipov14,Pouranvari/Yang/Seidel15}  
were studied, as well as inhomogeneous 
systems with random \cite{Laflo05,Igloietal07,Igloi/Lin08,Hoyosetal11}, 
aperiodic \cite{Igloi/Juhasz/Zimboras07} and exponentially decaying 
\cite{Vitagliano/Riera/Latorre10,Ramirez/Rodriguez/Sierra14}
couplings or random site energies \cite{Berkovits12,Pastur/Slavin14,Pouranvari/Zhang/Yang14}. 
Finally, systems in external potentials which
produce a varying density and surface regions have been investigated 
\cite{Eisler/Igloi/Peschel09,CV10a,CV10b,CV10c,CMV12a,Vicari12,
Eisler13,Eisler/Peschel14}.

In the present work, we look at yet another situation, namely at systems formed from
two pieces with different properties. Specifically, we study chains composed either of two 
critical parts, or a critical and a non-critial one. Such systems have been considered 
previously in the context of conformal invariance \cite{Hinrichsen90,Berche/Turban90,
Zhang/Li/Zhao96,Zhang/Chen/Li99}. 
In our case, they are realized in the form of undimerized or dimerized tight-binding models 
coupled by 
a special bond. Physically, this corresponds to either two metals with different bandwidths, 
or a metal and an insulator, and we will use this terminology in the following. In both cases 
one has two types of single-particle eigenfunctions: those essentially confined to one of the 
subsystems, and other ones extending through the whole system but having different wavelengths 
in the two parts. This is the same situation as for a potential step in quantum mechanics.
The occupied single-particle eigenfunctions determine the entanglement, and we study it  
between the two different pieces of the composite system in its ground state. 

For the metal-metal system at half filling, the asymptotic result is very simple. It turns 
out that only the interface bond matters and one comes back to the defect problem solved 
previously in \cite{Eisler/Peschel10}.
Thus $S$ varies logarithmically and the coefficient $c_{\mathrm{eff}}$ is determined 
by the transmission through the interface. The subleading terms, on the other hand, depend on 
the asymmetry but one can take this into account by a rescaling of the length and find links 
to conformal formulae based on the nature of the extended states.
Away from half filling, the entanglement is small as long as only the localized
states are occupied. It increases, when the extended states come into play, but there are
strong variations with the filling which depend on the size and are also connected with the ratio 
of the bandwidths, which brings an additional length scale into the problem. 

For the metal-insulator system at half filling, the entanglement lies systematically between 
that of the two pure systems. At large sizes, the insulator with its band gap dominates and 
limits the increase of $S$. For sizes smaller than the correlation length, however, there are
no states in the gap and one observes a logarithmic increase of the entropy. The system is
also a good case to compare the entanglement Hamiltonians on both sides. They have the same 
spectra, but their single-particle eigenfunctions and their explicit forms as hopping models
are quite different. Basically, one finds that the features are similar to those
of the pure systems on the corresponding side of the chain. For the eigenfunctions this means 
a decay from the interface into the interior which is slow in the metal and exponential in the 
insulator \cite{review09,Eisler/Peschel13}.

For the metal-metal system we also study the behaviour of the entanglement entropy after a
local quench in which the two pieces in their ground states are put together. Here the two 
Fermi velocities can be seen directly in the time structure and the result can be interpreted 
in the well-known picture of two emitted particles \cite{CC05}. 
 
The paper is organized as follows. In section 2 we describe the set-up und give the basic
formulae. In section 3 we investigate the metal-metal case by forming and diagonalizing
numerically the correlation matrix. We show the entanglement entropy and discuss its scaling
behaviour and filling dependence together with some entanglement spectra. In section 4 we 
consider the metal-insulator case at half filling and compare all relevant quantities with those 
of the pure systems. In section 5 we present the time evolution of $S$ after connecting two
metallic systems and Section 6 contains a summary. Finally, some analytical results for the
metal-metal system are given in an appendix.

\section{Setting and basic formulae}

 We consider open chains of free fermions with nearest-neighbour hopping and $2L$ sites. The
hopping is different in the left and right half and also between both parts. 
The Hamiltonian is
\begin{equation}
H= -\frac 1 2 \sum_{n=-L+1}^{L-1} \hat{t}_n \,(c^{\dag}_n c_{n+1} + c^{\dag}_{n+1} c_{n})
\label{ham}
\end{equation}

For the metal-metal system
\begin{equation}
 \hat{t}_n = \left\{ \begin{array}{r@{\quad:\quad}l}
                t_1 & n < 0 \\
                t_0 & n = 0  \\ 
                t_2 & n > 0
                \end{array} \right.  
\label{hopping_mm}
\end{equation}
and one has three parameters in the problem. However, by a rescaling of $H$, one can 
always achieve $t_1t_2=1$. We will assume this in the following and write the quantities as 
$t_1= \exp(\Delta)$, $t_2= \exp(-\Delta)$ and $t_0= \exp(\Delta_0)$. Moreover, we will
always consider positive $\Delta$, i.e. $t_1 > t_2$. In some places we also use the
ratio $r=t_2/t_1$.

\begin{figure}[thb]
\centering
\includegraphics[width=\textwidth]{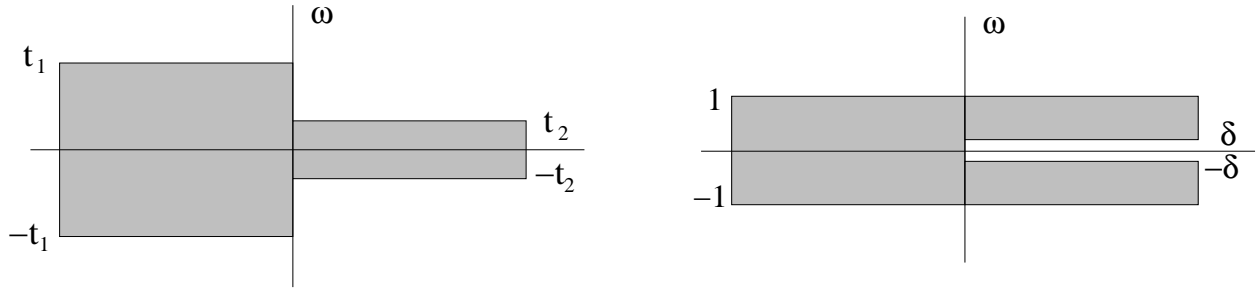}
\caption{Schematic band structure of the uncoupled composite systems. Left: Metal-Metal,
    Right: Metal-Insulator.}
\label{fig:bands}
\end{figure}

For vanishing coupling ($t_0=0$), the two parts of the chain have single-particle energies
\begin{equation}
\omega_{\alpha}=-t_{\alpha}\cos(q_{\alpha}),   \quad \alpha=1,2
\label{omega_met}
\end{equation}
with momenta $q_{\alpha}=\pi m/(L+1),\, m=1,2,...L$. The corresponding band structure with
bands between $\pm \,t_{\alpha}$ is shown on the left of Fig. \ref{fig:bands}. 
In the coupled system, one has
extended states for $|\omega| < t_2$, while outside this region they are confined essentially 
to the left half-chain. Moreover, one can have two states localized at the interface if 
$t_0$ is large enough.  

For the metal-insulator system we take 
\begin{equation}
 \hat{t}_n = \left\{ \begin{array}{r@{\quad:\quad}l}
                1  & n < 0 \\
                t_0 & n = 0  \\ 
                1+(-1)^{n+1} \delta & n > 0 
                \end{array} \right.  
\label{hopping_mi}
\end{equation}
In the insulator, one thus has alternating hopping, two sites per unit cell and the 
single-particle energies  
\begin{equation}
\omega_2= \pm \sqrt{\cos^2(q_2)+\delta^2\sin^2(q_2)}   
\label{omega_ins}
\end{equation}
where the momenta have to be determined from the boundary condition. Thus there is a gap
between $\pm \delta$. The resulting band structure is shown on the right of Fig. \ref{fig:bands}.
For an open chain, there are also states localized at the boundary with exponentially small $\omega_2$,
if the outermost bond is a weak one. The coupled system has extended states for $|\omega| > \delta$
and others, confined essentially to the metal, inside the gap. In this sense, it is still critical. 

%
%

It is simple to set up the eigenvalue problem for the full systems, and some details are
given in the Appendix. However, the matching
conditions in the center are difficult to handle in general. Therefore we determine
the single-particle eigenvalues $\omega_q$ and the corresponding (real) eigenfunctions
 $\Phi_q(n)$ numerically. The correlation matrix $C_{mn}=\langle c^{\dag}_m c_{n} \rangle$ 
is then  
\begin{equation}
C_{mn}=\sum_{q \in F} \Phi_q(m)\Phi_q(n) 
\label{corr}
\end{equation}
where the sum extends over the states $q$ in the Fermi sea.
Restricting $C_{mn}$ to the chosen subsystem (either the left or the right half-chain), 
its eigenvalues $\zeta_k$ give the single-particle 
eigenvalues $\varepsilon_k = \ln [({1-\zeta_k})/{\zeta_k}]$
of the free-fermion entanglement Hamiltonian $\mathcal{H}$ in
\begin{equation}
\rho = \frac{1}{Z} \exp(-\mathcal{H})
\label{rho}
\end{equation}
where $\rho$ is the reduced density matrix \cite{Peschel03}. From them, the entanglement 
entropy $S$ follows as
\begin{equation}
S =  \sum_k \ln (1 + \mathrm{e}^{-\varepsilon_k})+\sum_k
\frac{\varepsilon_k}{\mathrm{e}^{\varepsilon_k} +1}
\label{entropy}
\end{equation}
In addition to the eigenvalues $\zeta_k$ resp. $\varepsilon_k$, we also determine the 
corresponding eigenfunctions $\varphi_k$ and construct $\mathcal{H}$ in section 4.

\section{Metal-metal system}

\begin{figure}[thb]
\centering
\includegraphics[width=0.49\textwidth] {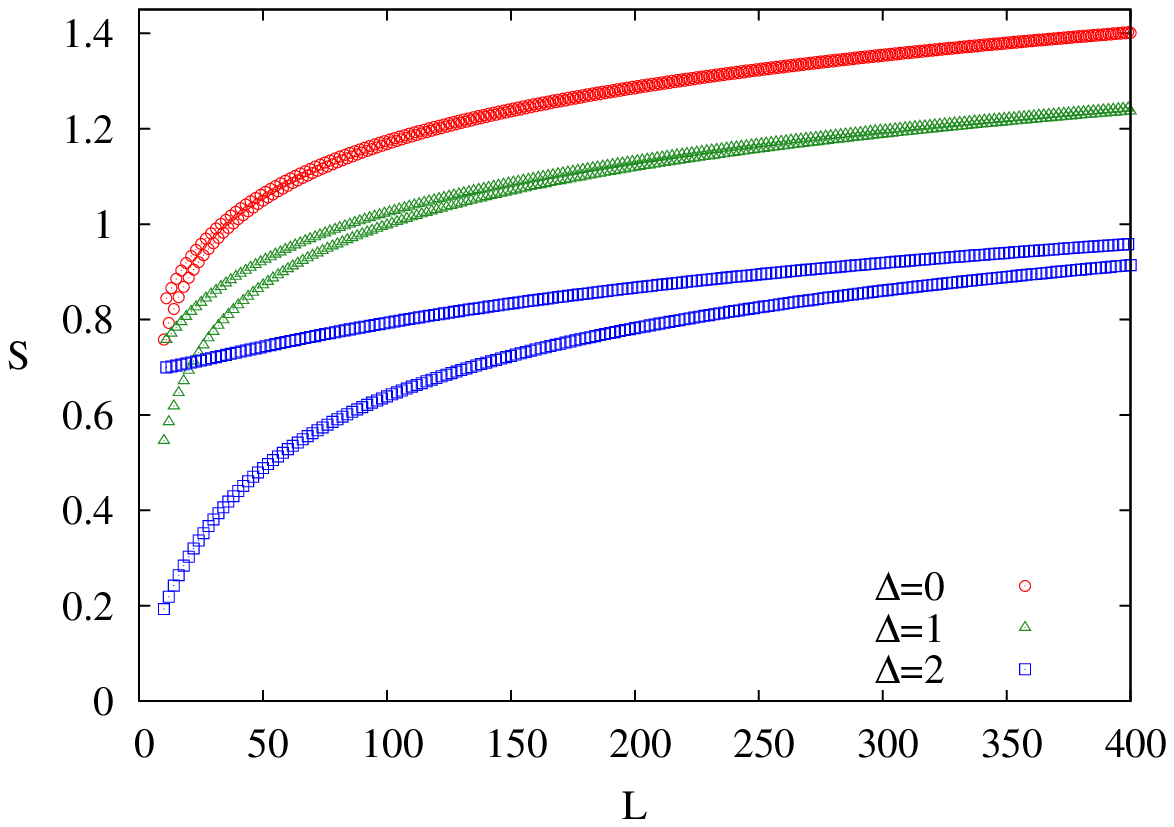}
\includegraphics[width=0.49\textwidth] {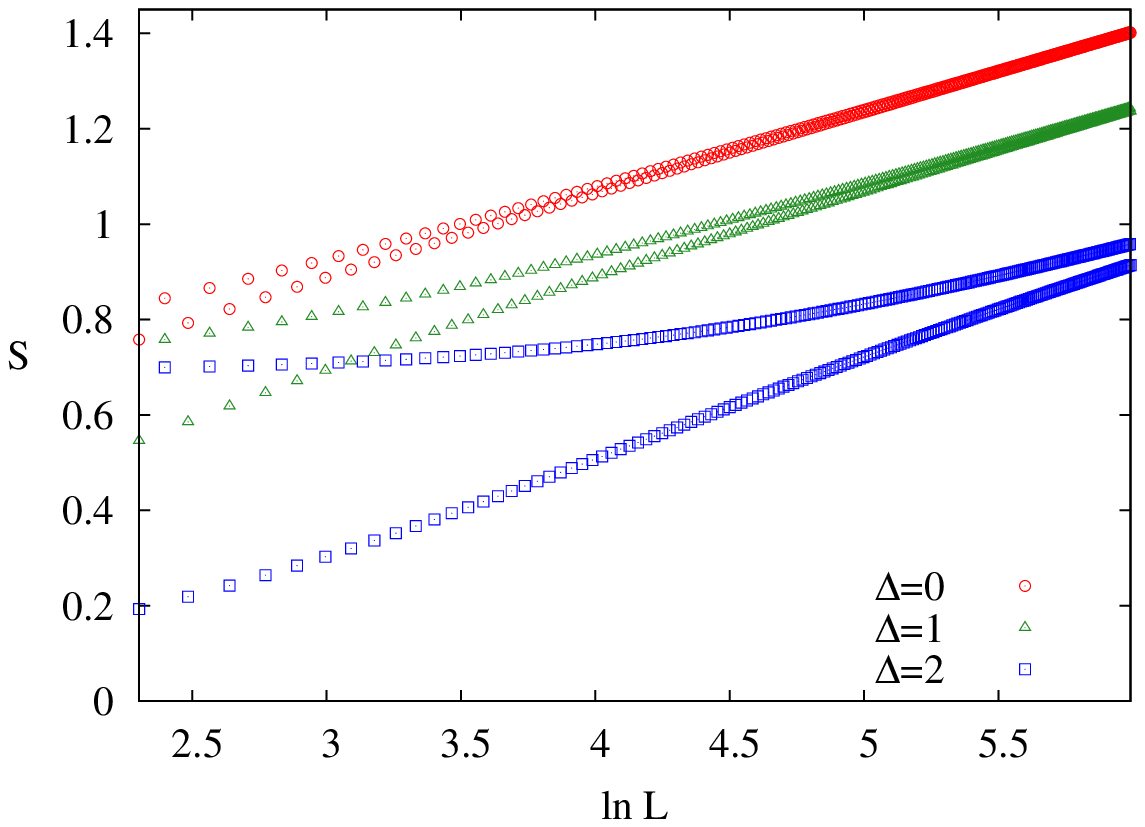}
\caption{Entanglement entropy $S$ for $\Delta_0=0$ and three values of $\Delta$. Left: as
   function of $L$. Right: as function of $\ln L$. The upper (lower) curves correspond to
   odd (even) values of $L$. }
\label{ent_mm_Delta}
\end{figure}

We first consider half-filled systems where the Fermi level is in the middle of the bands.
The resulting entanglement entropies for $t_0=1$ and three values of the parameter $\Delta$ 
are shown in Fig. \ref{ent_mm_Delta}, both for even and for odd $L$. One sees that the typical increase with 
the size, known for the homogeneous case $\Delta=0$, persists in the composite systems. The
plot against lnL shows that also the asymptotic law $1/6\ln L+k$ is unchanged. However, the 
value of $k$ becomes smaller and the finite-size effects, in particular the even-odd 
alternation, increase dramatically as $\Delta$ becomes larger.
For $\Delta=3$ (not shown), the values of $S$ are initially close to zero and to $\ln2$, 
respectively.
In this case, the ratio of the bandwidths $t_2/t_1 \simeq 1/400$ is already very small, the 
states localized on the left drop rapidly on the right, while the extended states have very 
small amplitudes on the left (see the Appendix). For even $L$, this leads to a small 
entanglement which only increases as more states come into play at larger $L$. The value 
$\ln2$ for odd $L$ is a consequence of the particle-hole symmetry of the problem which, in
this case, forces one of the $\varepsilon_k$ to vanish. 
%
%

\begin{figure}[thb]
\centering
\includegraphics[width=0.49\textwidth]{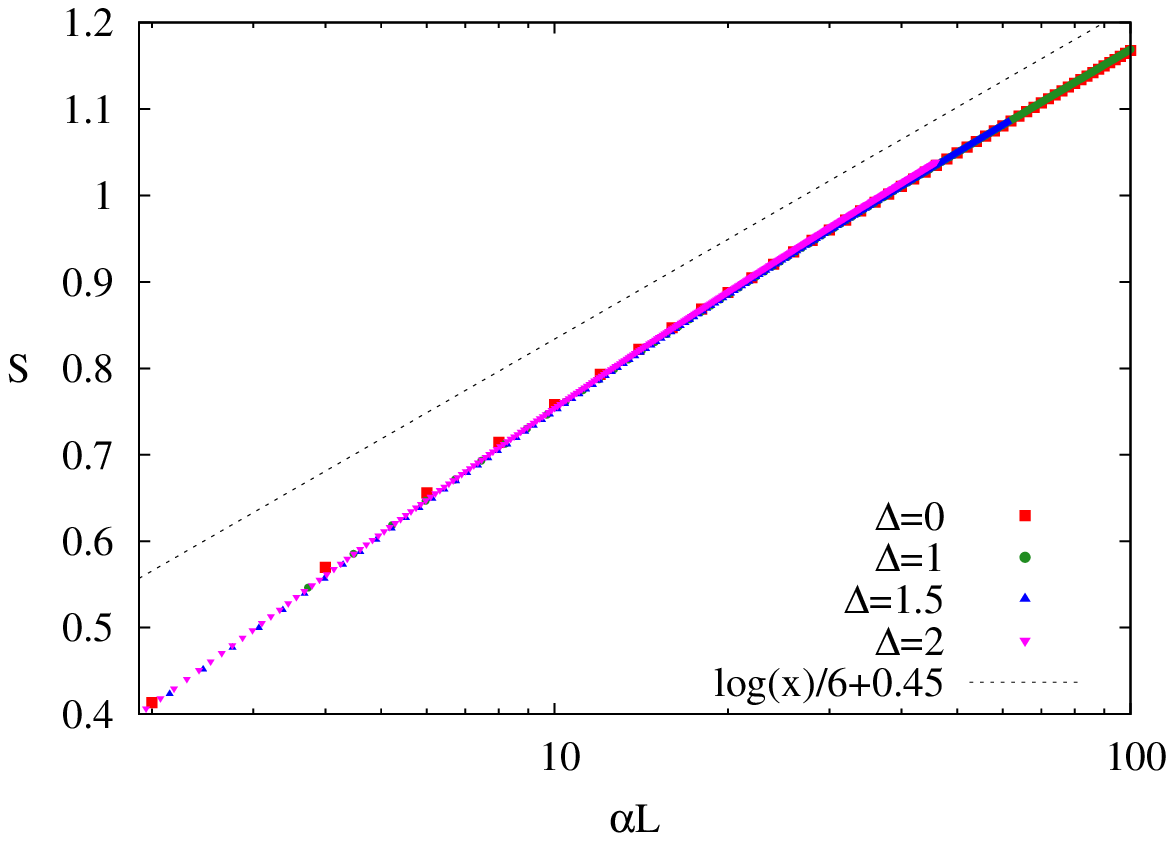}
\includegraphics[width=0.49\textwidth]{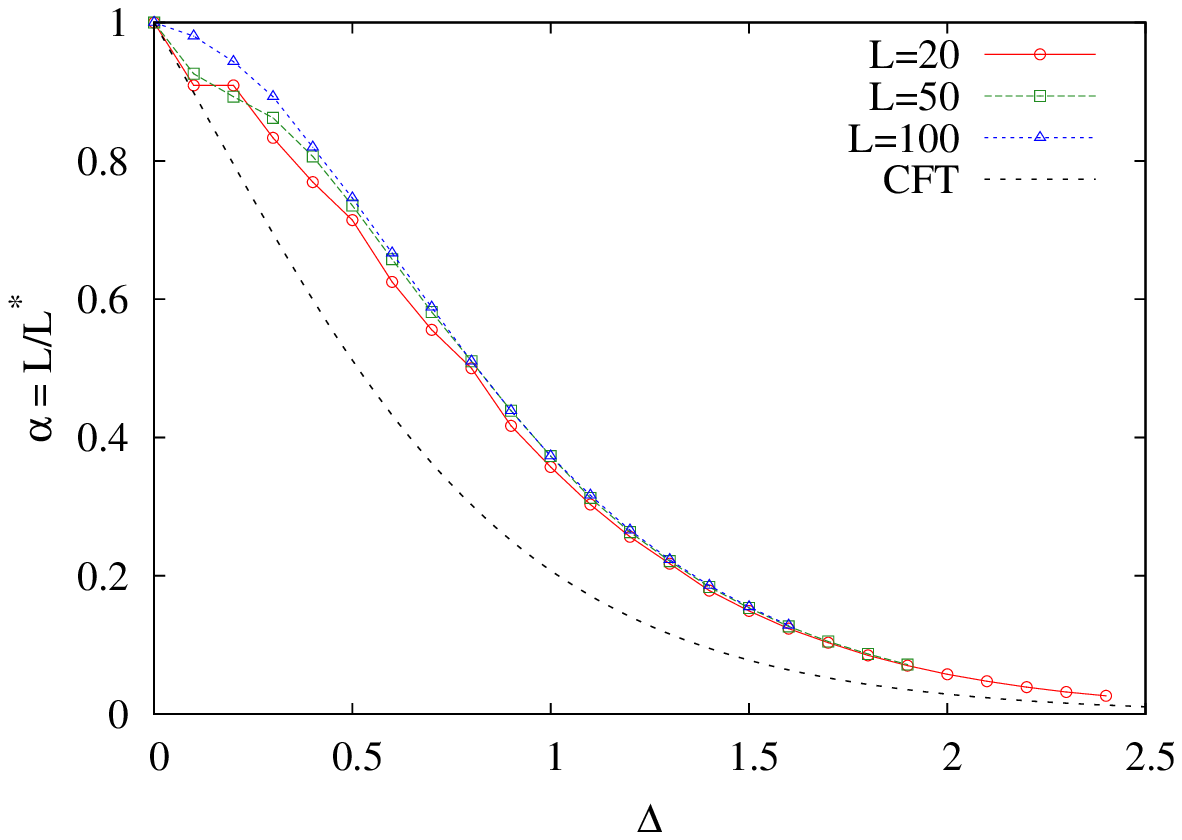}
\caption{Scaling behaviour of the entanglement entropy. Left: $S$ as a function of the effective 
  length $\alpha L$ for four values of $\Delta$ (logarithmic scale). Right: Variation of 
  $L/L^*=\alpha$ with $\Delta$. The dotted line is the conformal factor in (\ref{ent_FC}).}
\label{ent_mm_scaling}
\end{figure}

The function $S(L,\Delta)$ has an intriguing universal behaviour. If one defines a length
$L^*= L/\alpha$ and chooses the scale factor $\alpha$ properly, one can achieve
$S(L^*,\Delta)= S(L,0)$, i.e. all curves collapse on the one for the homogeneous
system. Written differently, one has 
\begin{equation}
S(L,\Delta)=S(\alpha L,0)
\label{scaling}
\end{equation}
In the asymptotic region, where one has straight lines in a logarithmic plot, this feature
is not surprising: A rescaling can always generate the necessary shift to make the
curves coincide. However, the phenomenon is not restricted to this region. As shown in 
Fig. \ref{ent_mm_scaling} on the left, it also holds for small sizes where there is still curvature in the graph.
The variation of $\alpha$ with $\Delta$ is given on the right hand side of Fig. \ref{ent_mm_scaling}. One 
sees that the results determined from different (even) $L$ agree very well and that
$\alpha$ becomes rapidly smaller as $\Delta$ increases. The curve can be described approximately
by $1/\ch^2 \, \Delta$. Note that the relation (\ref{scaling}) can only be applied to values
of $L$ such that $\alpha L$ is larger than 1. 

%
%

At this point, an observation from the appendix is helpful. Namely, near the Fermi level the 
wavefunctions in the right subsystem are the same as for a homogeneous system of total length 
$L(1+r)$. Thus, ignoring the states outside the narrow band, the correlation matrix on the right 
is the same as for an unsymmetrically divided homogenous chain and one can invoke the conformal 
result for the entropy \cite{CC04} extended in \cite{Fagotti/Calabrese11} to include $1/L$ 
corrections.
This gives 
\begin{equation}
S_{FC} = \frac{1}{6} \ln z - \frac{(-1)^{\ell}}{z} + \mathrm{const.} , \quad \quad  
z=\frac{4}{\pi}L(1+r)\sin{\frac{\pi \ell}{L(1+r)}},
\label{ent_FC}
\end{equation}
with $\ell=L$.
The $r$-dependent factor in $z$ provides a similar rescaling of $L$ as $\alpha$, and a corresponding
shift of the curves. It turns out, however, that this is not enough to make them coincide. Rather, 
one has a residual difference $S-S_{FC}=S_I(\Delta)$, which one can attribute to the interface, and
which rises smoothly from $0$ to about $0.13$ as $\Delta$ increses from $0$ to $3$. It has its 
origin in the neglected exponentially decaying states below the narrow band. 

\begin{figure}[thb]
\centering
\psfrag{l}[][][.9]{$\ell$}
\includegraphics[scale=.8]{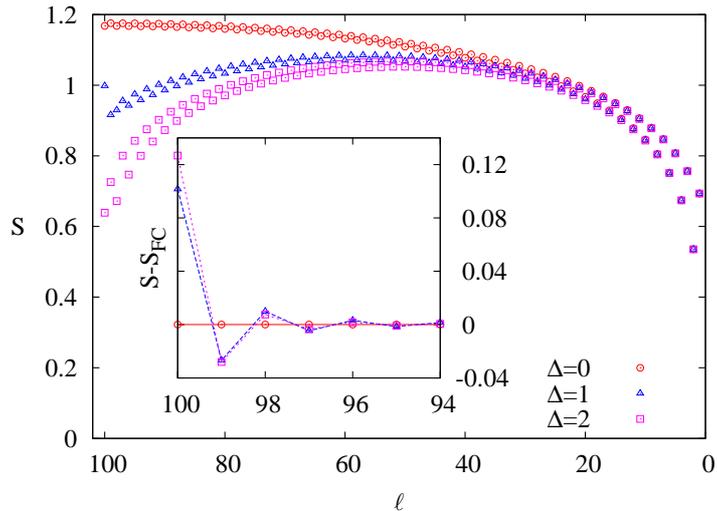}
\caption{Entanglement entropy between the outermost $\ell$ sites of the right subsystem and the remainder
     for $L=100$ and three values of $\Delta$. The inset shows the deviation from the formula (\ref{ent_FC}).}
\label{ent_mm_scaling2}
\end{figure}

This can be seen clearly, if one considers a partition of the system with $\ell < L$ sites on the 
right and calculates the entanglement between this subsystem and the remainder. As demonstrated in
Fig. \ref{ent_mm_scaling2}, the expression (\ref{ent_FC}) then fits the data very well as soon as one moves away 
from the interface by a few lattice sites. One can also consider a division located in the 
left part of the system where the extended wave functions are the same as in a homogeneous system 
with total length $L(1+1/r)$, but in this case the conformal formula does not work well 
because the additional states are more important. These results also show that the 
symmetrical division is actually a somewhat special case.
%
%

\begin{figure}[thb]
\centering
\includegraphics[width=0.49\textwidth]{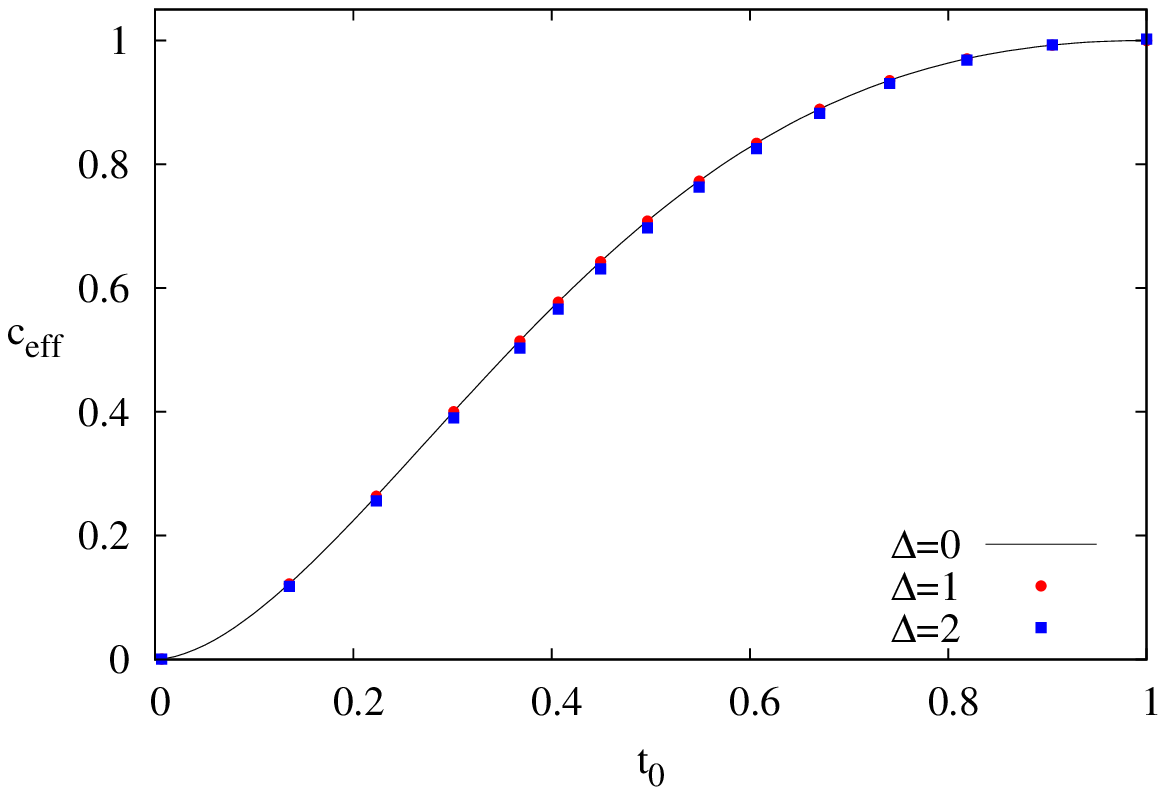}
\includegraphics[width=0.49\textwidth]{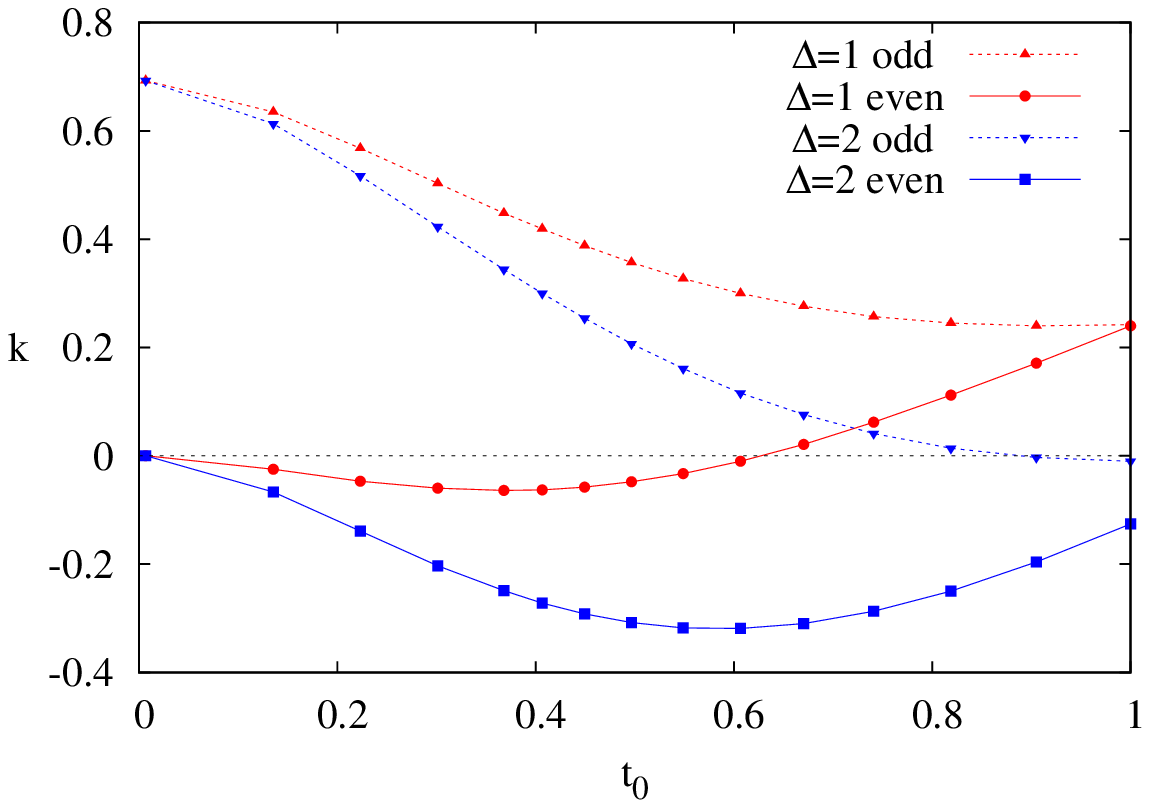}
\caption{Effective central charge $c_{\mathrm{eff}}$ and constant $k$ in (\ref{c_eff}) as functions
      of $t_0$ for different values of $\Delta$. The line $\Delta=0$ in the left figure 
      is the theoretical result, see \cite{Eisler/Peschel10}.}
\label{ceff_mm}
\end{figure}

We now turn to the effect of $t_0$. If one varies the central bond in a homogeneous chain 
($\Delta=0$), the asymptotic behaviour of $S$ is
\begin{equation}
S = \frac{c_{\mathrm{eff}}}{6} \ln L + k
\label{c_eff}
\end{equation}
where $c_{\mathrm{eff}}$ depends on $t_0$ and is given by an explicit, though lengthy formula in which 
only the transmission coefficient at the Fermi energy $T \equiv s^2$ enters \cite{Eisler/Peschel10}.
But the calculation in the Appendix shows that $T$ is independent of $\Delta$ and given by
\begin{equation}
T = \frac{1}{\ch^2 \, \Delta_0}
\label{transmission_center_text} 
\end{equation}
Therefore one expects the result of the homogeneous problem also in the composite 
systems. This is, in fact, the case and demonstrated in Fig. \ref{ceff_mm} on the left, where results 
obtained by fitting the data between $L=100$ and $L=400$ to (\ref{c_eff}) plus a $1/L$ term 
are collected. For $\Delta=2$, the average between even and odd sizes was taken, while for 
$\Delta=1$ this was unnecessary. This verifies the Fermi-edge character of $c_{\mathrm{eff}}$ also
in the present case.     
%
%

The constant $k$, on the other hand, depends on $\Delta$, as was found above already for
the special case $t_0=1$. It is shown in Fig. \ref{ceff_mm} on the right. The values for odd $L$
are always larger than those for even $L$ with the maximal difference of $\ln2$ appearing for 
$t_0 \rightarrow 0$. This is connected with the half-integer particle number in each
subsystem for odd $L$ which leads to a kind of singlet state even in the decoupling limit. 
The minimum in $k$ for even $L$ is also present for $\Delta=0$ and was also found 
for a segment in an infinite chain \cite{Peschel05}. 

\begin{figure}[thb]
\centering
\includegraphics[scale=.8]{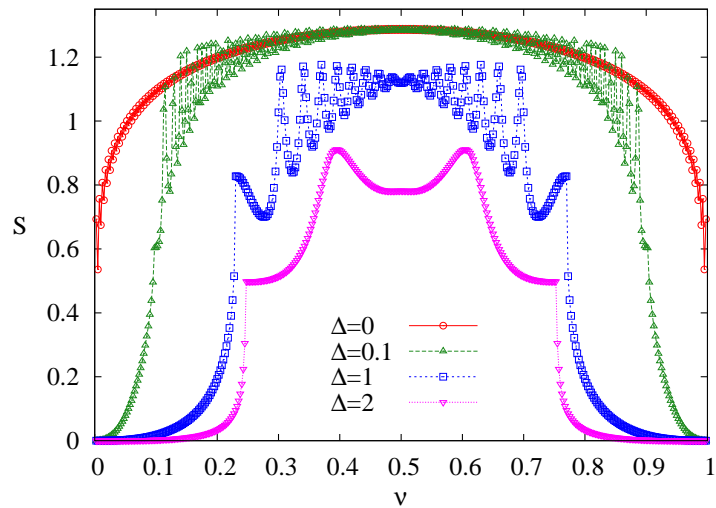}
\caption{Entanglement entropy $S$ as a function of the filling $\nu=N/2L$ for $L=200$
     and different $\Delta$.}
\label{ent_mm_filling}
\end{figure}

Finally, we consider the dependence of $S$ on the filling $\nu=N/2L$ of the system, where $N$ 
is the number of particles. It is shown in Fig. \ref{ent_mm_filling} for a fixed length $L=200$ and four values 
of $\Delta$. The smooth curve (red) is the result for the homogeneous system and given, up to 
the small even-odd oscillations \cite{Laflorencie/etal06,Affleck/Laflorencie/Sorensen09,
Calabrese/Essler10,Fagotti/Calabrese11} and a constant, by 
$1/6 \ln(\sin{q_F})$, where $q_F=\pi \nu$ \cite{Jin/Korepin04}. Turning on $\Delta$, regions of 
very small entanglement develop for small
and large filling, where only states localized on the left are occupied or remain empty.
At the same time, the oscillations of $S$ in the central region become slower and slower. 
%
%

\begin{figure}[thb]
\centering
\includegraphics[width=0.49\textwidth]{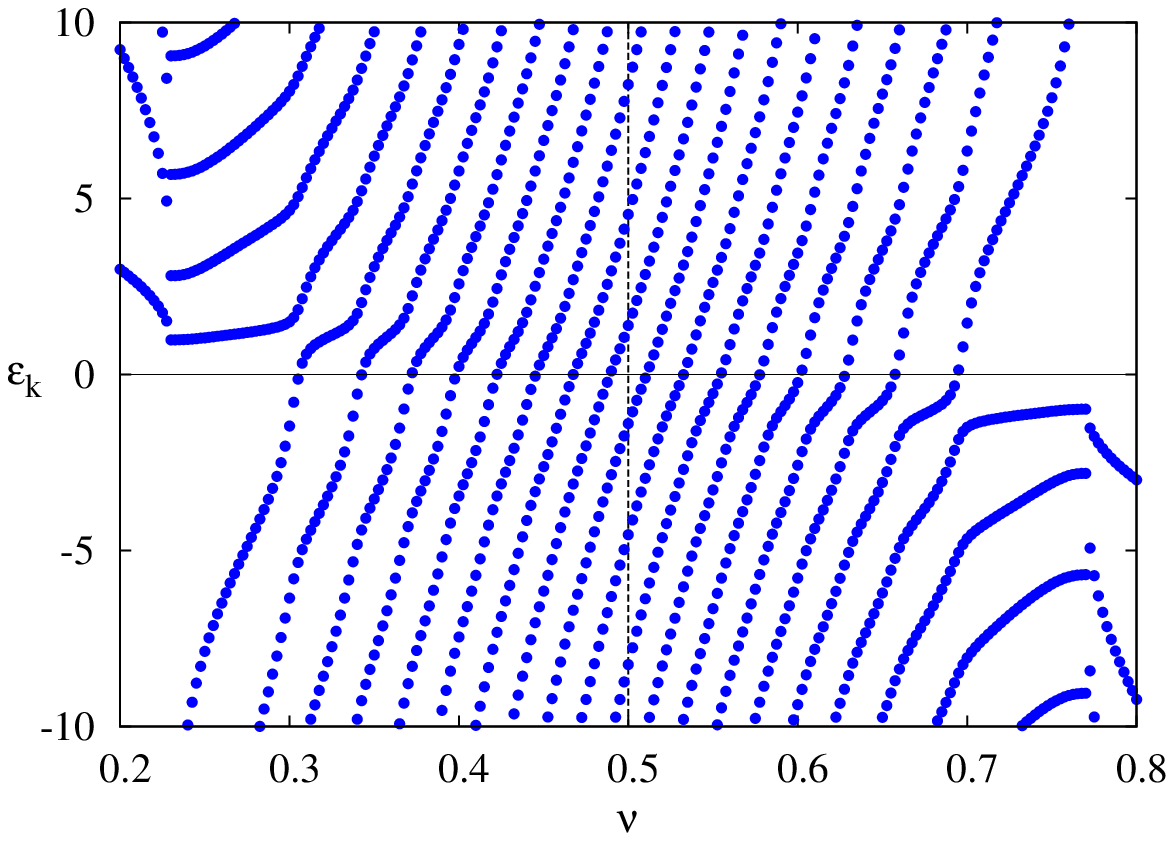}
\includegraphics[width=0.49\textwidth]{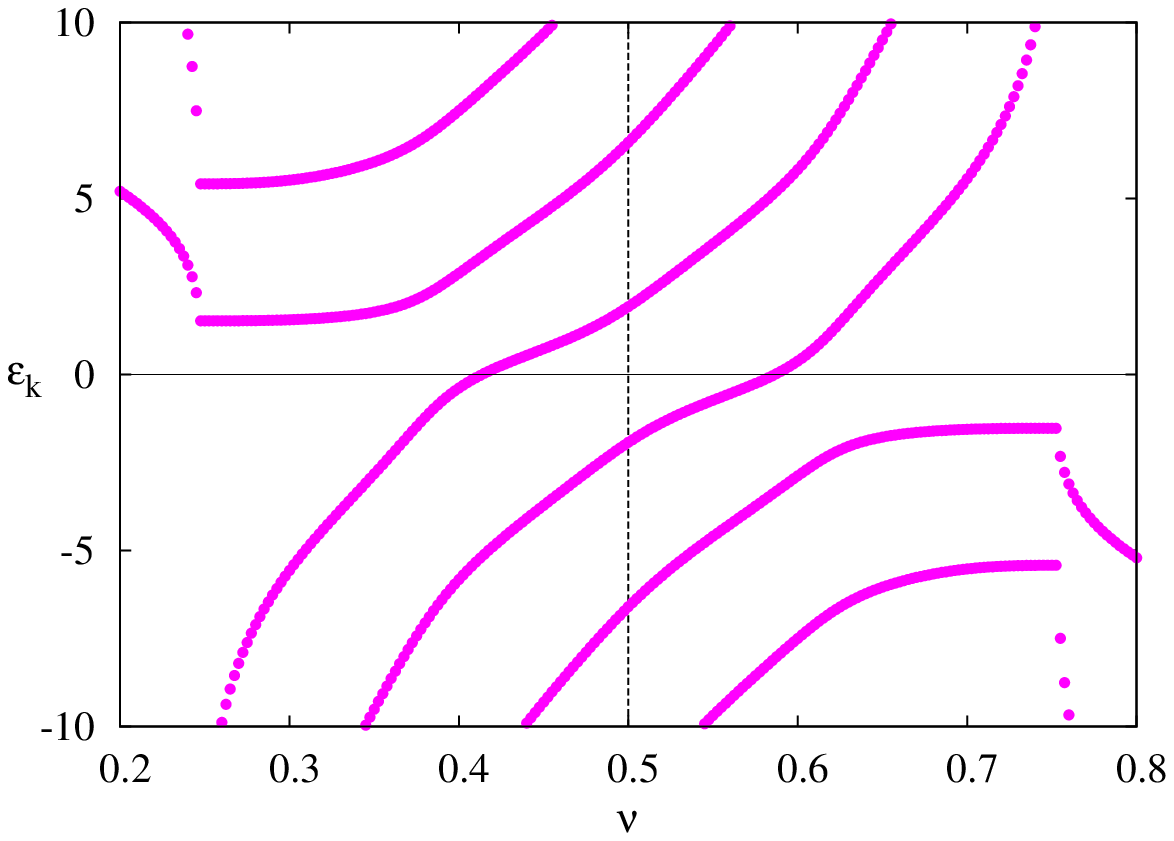}
\caption{Variation of the low-lying single-particle eigenvalues $\varepsilon_k$ with the filling 
     $\nu=N/2L$ for $L=200$. Left: $\Delta=1$. Right: $\Delta=2$. The colours are the same as 
     in Fig. \ref{ent_mm_filling}.}
\label{eps_mm_filling}
\end{figure}

The origin of these oscillations, at the level of the eigenvalues $\varepsilon_k$, can be
seen in Fig. \ref{eps_mm_filling}. There, the low-lying $\varepsilon_k$ are shown for all fillings
between $0.2$ and $0.8$. For the case $\Delta=2$, shown on the right, the variation of the
$\varepsilon_k$ with $\nu$ is rather slow and the curves cross zero only at two fillings.
At these points, the crossing eigenvalue gives the maximal contribution  to $S$, namely $\ln2$,
and they coincide with the locations of the maxima in Fig. \ref{ent_mm_filling}. For 
$\Delta=1$, shown on the left, the variation is much faster and one has altogether 16 
crossings. This is again the number of maxima of $S$ in the center of Fig. \ref{ent_mm_filling}.
%
%

These structures can again be understood from the relation to a smaller, unsymmetrically
divided homogeneous system. In such a case, if $L'$ is the size of the smaller subsystem, 
one has $L'$ crossings of the eigenvalues and the effect on $S$ is given by a factor 
$\sin{(q_F(2L'+1))}/\sin{q_F}$ replacing the alternating sign of the $1/z$ term in (\ref{ent_FC}). 
This leads to slow oscillations, although less pronounced than the observed ones. However, 
$L'=rL$ gives $L'=3.6$ for $\Delta=2$ and $L'=27$ for $\Delta=1$ and thus not the correct numbers 
(2 and 16), 
even if one takes the nearest integer. But one can argue that the length $rL$ only refers to the 
states in the center of the band, while near the edges $\sin{k_2}$ appears in (\ref{momenta3}) 
which is smaller than $k_2$ by a factor of $2/\pi$. Working with $\tilde{L'}=2rL/\pi$ 
gives values $2.3$ and $17.2$, respectively, which are quite close to the numerical findings. 
One also arrives at $\tilde{L'}$ by counting the number of levels which the left subsystem 
contributes to the total number of band states. To accommodate them, one needs just this number 
of additional sites.

One should mention that one can see basically the same structures also in the energy $\omega$ 
as tiny deviations from the otherwise smooth level spacing of the extended states. In this 
case, the interpretation is that $\tilde{L'}$ also gives the number of periods of the slowly 
varying tangent in (\ref{momenta3}) as $k_2$ sweeps through the band, and thus the number of 
possible peculiarities in the allowed momenta.

\section{Metal-insulator system}

\begin{figure}[thb]
\centering
\includegraphics[width=0.49\textwidth]{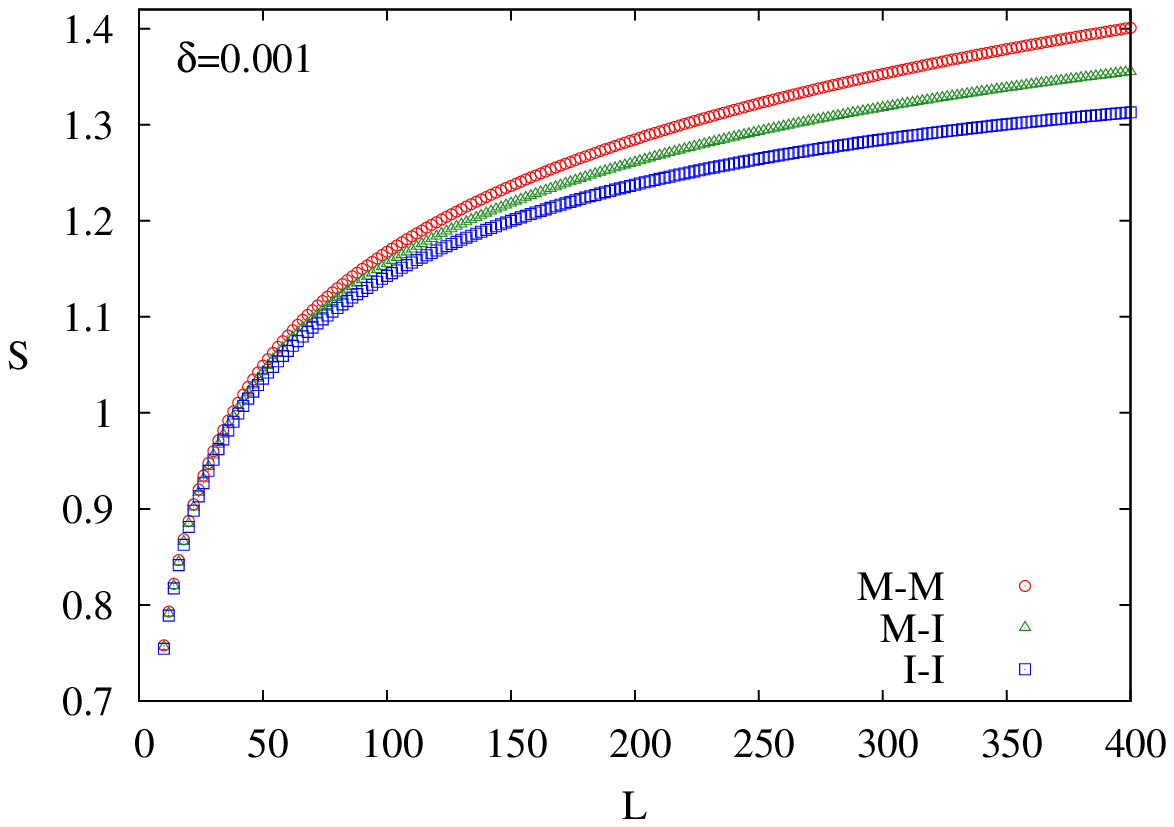}
\includegraphics[width=0.49\textwidth]{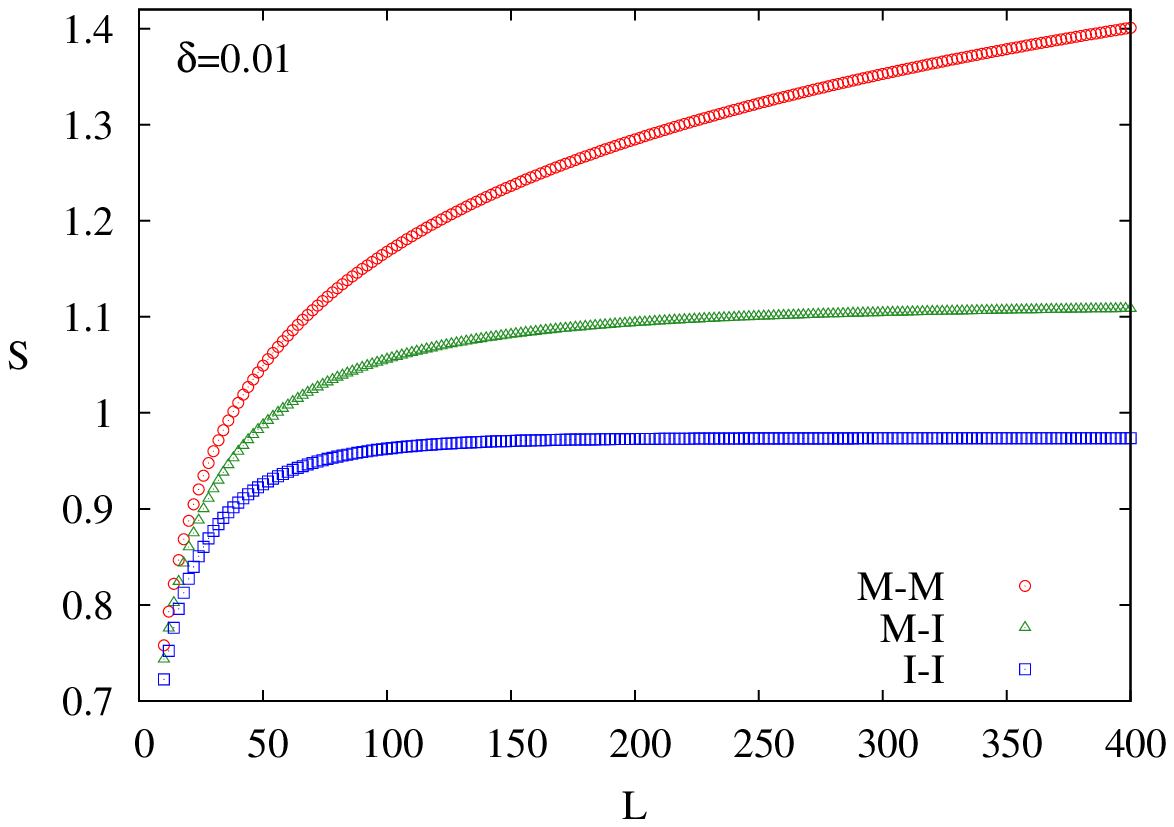}
\caption{Entanglement entropies for metal-metal, metal-insulator and insulator-insulator
systems as a function of $L$ for two values of the dimerization parameter $\delta$.}
\label{ent_mi}
\end{figure}

In the following, we always consider half-filled systems. To avoid boundary states in the 
dimerized subsystem, we work with even $L$
and choose strong bonds $1+\delta$ at the borders. The entanglement entropies
for two relatively small dimerizations and $t_0=1$ are shown in Fig. \ref{ent_mi}, together with those
for the pure systems. One sees that the mixed system always has an intermediate entanglement,
which is very plausible. Replacing one half of the metal by an insulator reduces the metallic 
result, while replacing one half of the insulator by a metal increases the insulator result.
%
%
For $\delta=0.001$, shown on the left, all three curves look similar and there is no
sign of a saturation, while for $\delta=0.01$ a saturation is clearly visible for the
lower ones. This can be interpreted in terms of the correlation length $\xi=1/\delta$
in the insulator. It is $\xi=1000$ in the first case so that the non-criticality  
does not fully show up. In the second case, where $\xi=100$, it does and the
crossover takes place roughly around this value.

In the region $L \ll \xi$, one can fit the curves for the composite system to a form
$a\ln L +b$ and finds values of $a$ quite close to $1/6$. This can also be done for $t_0<1$
and gives curves for $c_{\mathrm{eff}}$ which lie somewhat below the metallic one shown in 
Fig. \ref{ceff_mm}, but approach it as $\delta$ goes to zero. 
In the opposite limit, $L \gg \xi$, the entropy saturates and for $t_0=1$ one finds the asymptotic 
law $S \simeq 1/6 \ln\xi + \mathrm{const.}$ as for the pure insulator but with a larger constant.
To obtain the coefficient $1/6$ accurately from not too large $\xi$, one has to include subleading
corrections of the form $(d\ln\xi+e)/\xi$ which one knows in the pure case from the connection to 
a transverse Ising chain sketched below. Again, the analysis can be extended to $t_0<1$, and in
this case one finds $c_{\mathrm{eff}}$ to an accuracy of 3-4 digits.

On the level of the wavefunctions entering the correlation matrix, the
(pseudo-) critical behaviour can be understood in the following way. For $L < \xi$ one has
$|\omega| \sim 1/L > \delta$ and there are no states in the metal with energy in the
band gap of the insulator and none in the insulator too close to the band edge. Thus one
only has extended wavefunctions which are similar to those in the pure metal. This can be
seen analytically or from the numerics.

%
\begin{figure}[thb]
\centering
\includegraphics[scale=.7]{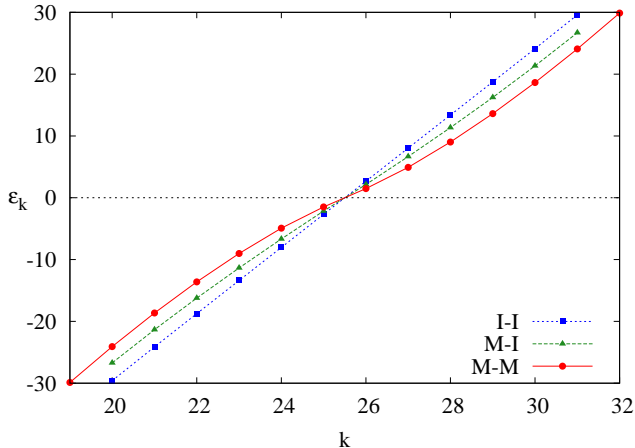}
\caption{Eigenvalues $\varepsilon_k$ for metal-metal, metal-insulator and insulator-insulator systems
with $L=50$ and $\delta=0.1$ for the insulator.}
\label{epsilon_mi}
\end{figure}
%

It is also instructive to look at the $\varepsilon_k$ spectra. In Fig. \ref{epsilon_mi} an example is
shown for the case $\delta=0.1$ and $L=50$. The lowest curve with the slight bend is the well-
known result for the metal, while the highest one is for the insulator and strictly linear
as for the transverse Ising chain or the XY model \cite{review09}. In fact, one can find a 
relation between the $\varepsilon_k$ for the dimerized hopping model and the transverse Ising 
chain using the method in \cite{Igloi/Juhasz08}. The latter model then has field $h=1$ and 
coupling $\lambda=(1-\delta)/(1+\delta)$ and the $\varepsilon_k$ are
\begin{equation}
\varepsilon_k = (2k+1)\,\varepsilon, \quad \quad k=0,\pm 1,\pm 2,... , \quad \quad
         \varepsilon=\pi\,I(\lambda')/I(\lambda)
\label{eps_dim}
\end{equation}
where $I(\lambda)$ denotes the complete elliptic integral of the first kind and 
$\lambda' = \sqrt{1-\lambda^2}$, see also \cite{Sirker14}.
The MI result lies in between, with a small break after the  first eigenvalue, 
which becomes more apparent for larger values of $\delta$, when the curve becomes steeper and
more linear.

\begin{figure}[thb]
\centering
\includegraphics[width=0.49\textwidth]{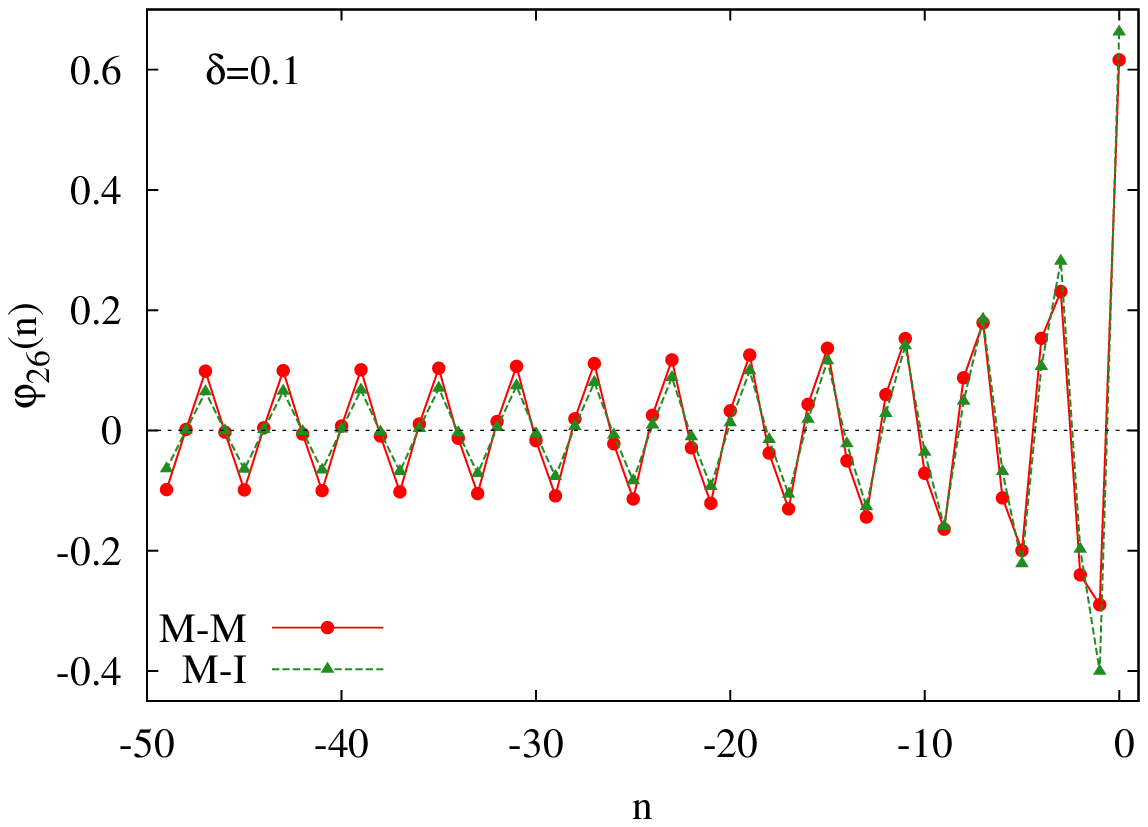}
\includegraphics[width=0.49\textwidth]{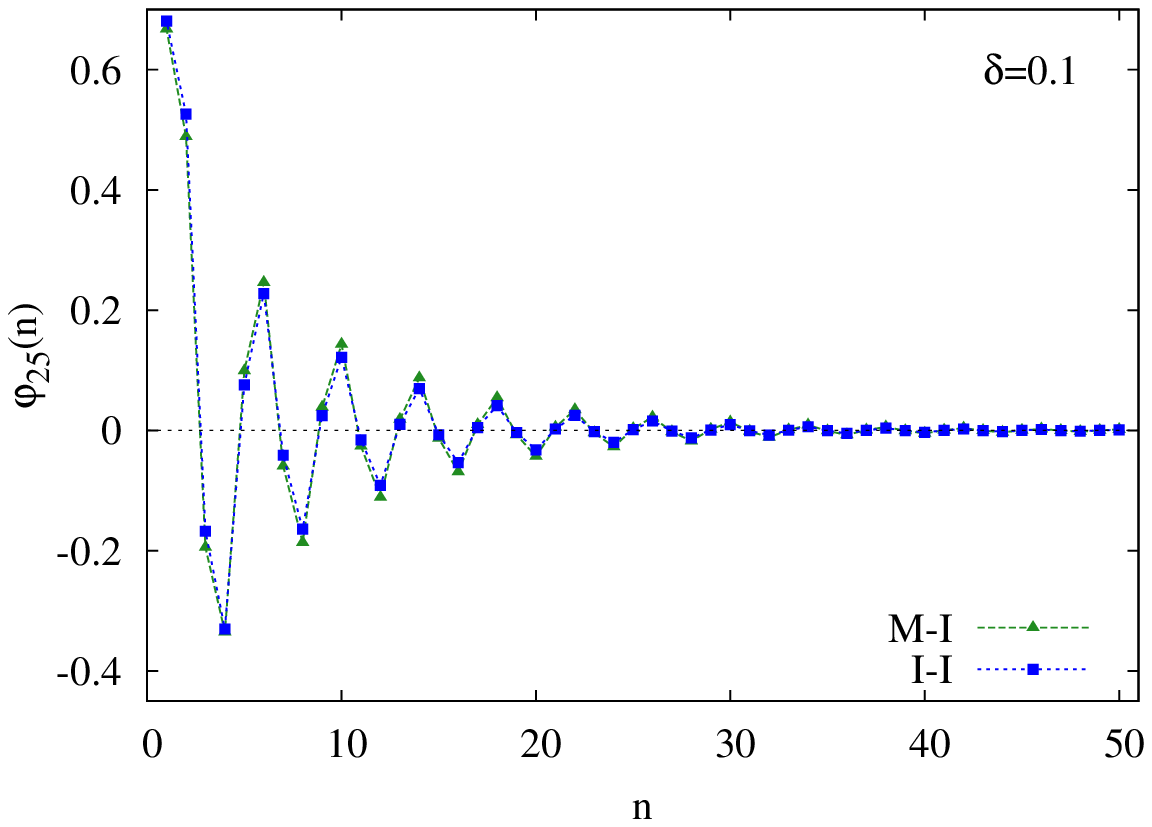}
\caption{Eigenvectors of the MI correlation matrix for the smallest $|\varepsilon_k|$ in the
metallic (left) and the insulating (right) subsystem for $L=50$ together with the results for the 
homogeneous cases (MM and II).}
\label{eigenvectors}
\end{figure}

For the spectrum, it does not matter in which of the two subsystems one considers the 
correlation matrix. Formally, this follows from the property $C^2=C$ for the full system 
\cite{Turner/Zhang/Vishwanath09}. However, the eigenvectors of the two reduced matrices,
and therefore the corresponding entanglement Hamiltonians $\mathcal{H}$, are quite different. 
This is illustrated in Fig. \ref{eigenvectors} for the lowest $|\varepsilon_k|$ and $\delta=0.1$.
We have  plotted $\varphi_k(n)$ for $+\varepsilon_k$ on the left and for $-\varepsilon_k$ on the
right, since these two are directly related. Namely, if one partitions $C$ into four blocks 
according to the location of the sites
and $\varphi^1$ is an eigenvector of $C^{11}$ for part 1 with eigenvalue $\zeta$, then 
\begin{equation}
\varphi^2 =  C^{21} \varphi^1
\label{left_right_ev}
\end{equation}
is an eigenvector of $C^{22}$ for part 2 with eigenvalue $1-\zeta$.

One sees the typical features of the pure systems 
on the two sides, namely a slow power-law decay in the metal (left) and a fast one within 
a distance $\xi=10$ in the insulator (right). The pure cases are shown for comparison, and 
one notes that the differences in the composite system are relatively small, lying more in 
the amplitudes than in the general behaviour. 

%
%

With the eigenvectors, one can explicitly determine the entanglement Hamiltonians $\mathcal{H}$.
They have again the form of hopping models with basically only nearest-neighbour hopping
as in $H$, but the amplitudes $\tilde{t}_n$ increase from the center towards the boundaries. 
\vspace{0.2cm}
%
\begin{figure}[thb]
\centering
\includegraphics[width=0.49\textwidth]{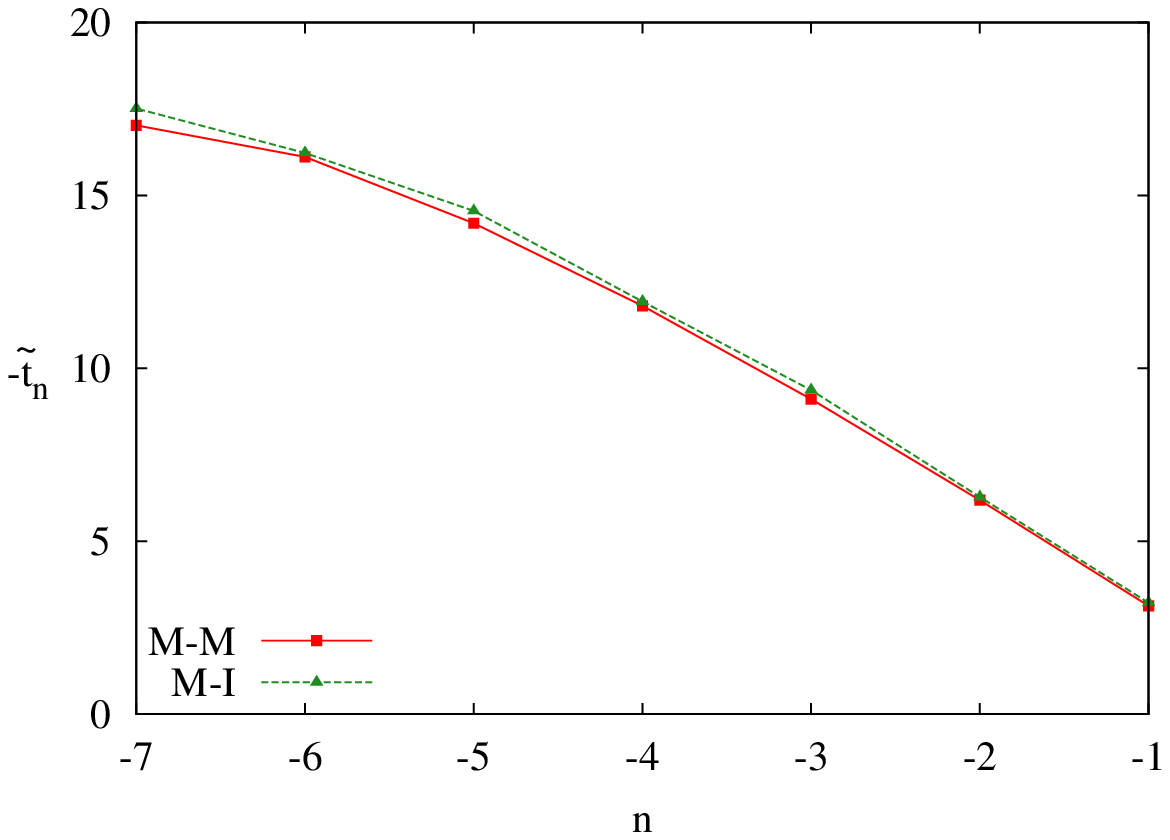}
\includegraphics[width=0.49\textwidth]{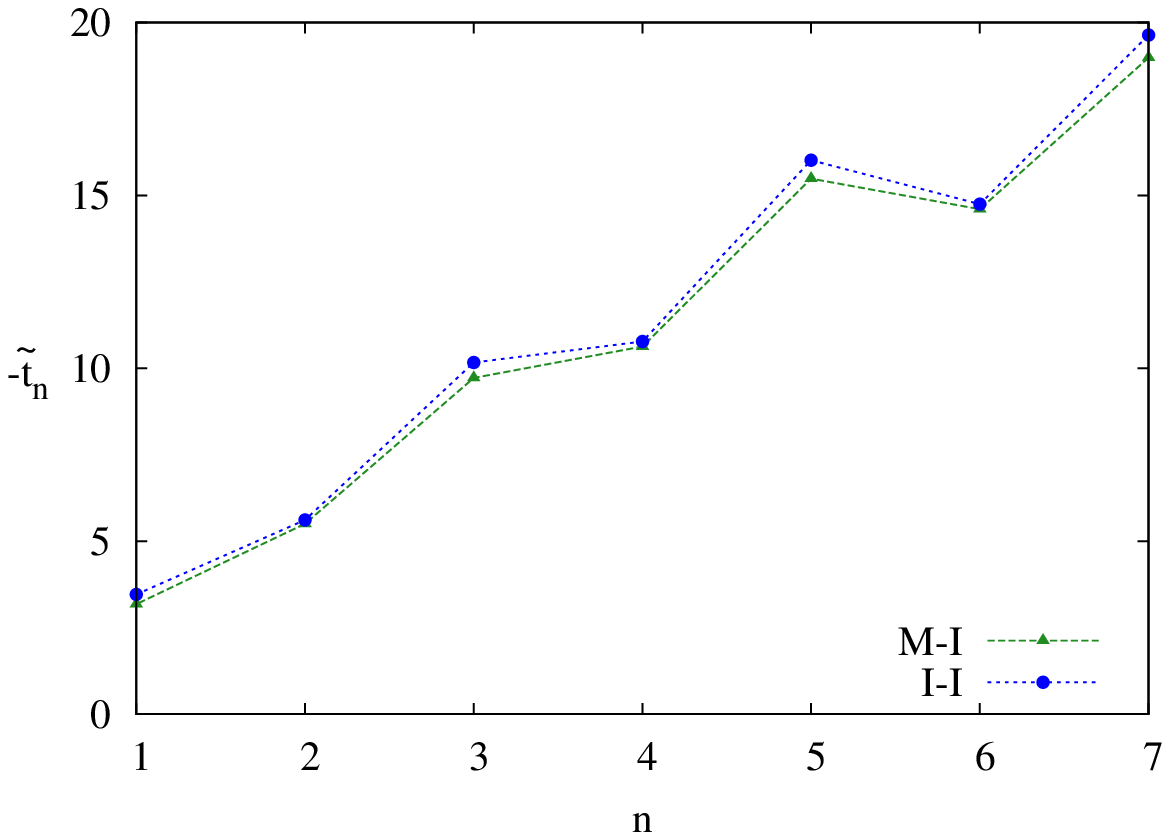}
\caption{Nearest-neighbour hopping in the entanglement Hamiltonians for the left and right
     subsystem as a function of the position together with the results for the pure systems
     for $L=8$ and $\delta=0.1$.}
\label{h_ent}
\end{figure}
%


In the pure metal, this increase is monotonous and the $\tilde{t}_n$ vary as $n(2L-n)$ 
\cite{review09}, while in the insulator one has an additional alternation which is coupled to
the bond alternation in $H$. Due to the form of 
the eigenvectors, the result in the composite system is again close to that of the pure system 
on the corresponding side. This is shown in Fig. \ref{h_ent} for a system with a total of $2L=16$ sites, 
which is the largest size attainable before numerical errors in the large $\varepsilon_k$ set in.

\pagebreak

\section{Evolution after a quench}

 In this section, we return to the metal-metal systems and study how the entanglement 
evolves after one joins the two initially disconnected parts. This type of local quench 
has been considered repeatedly in the past 
\cite{Eisler/Peschel07,CC07,EKPP07,Schoenhammer07,Klich/Levitov09,Stephan/Dubail11,
Igloi/Szatmari/Lin12,
Eisler/Peschel12,Collura/Calabrese13,Alba/HDM14,Kennes/Meden/Saleur14,Asplund/Bernamonti14,
Chen/Vidal14,Thomas/Flindt15}. For the homogeneous case, one finds 
oscillations of $S(t)$ which one could call ``entanglement bursts'', see e.g. 
\cite{Stephan/Dubail11}. 
In our case, the situation is more complex, because of the interface and the two Fermi 
velocities.

\begin{figure}[thb]
\centering
\includegraphics[width=0.49\textwidth]{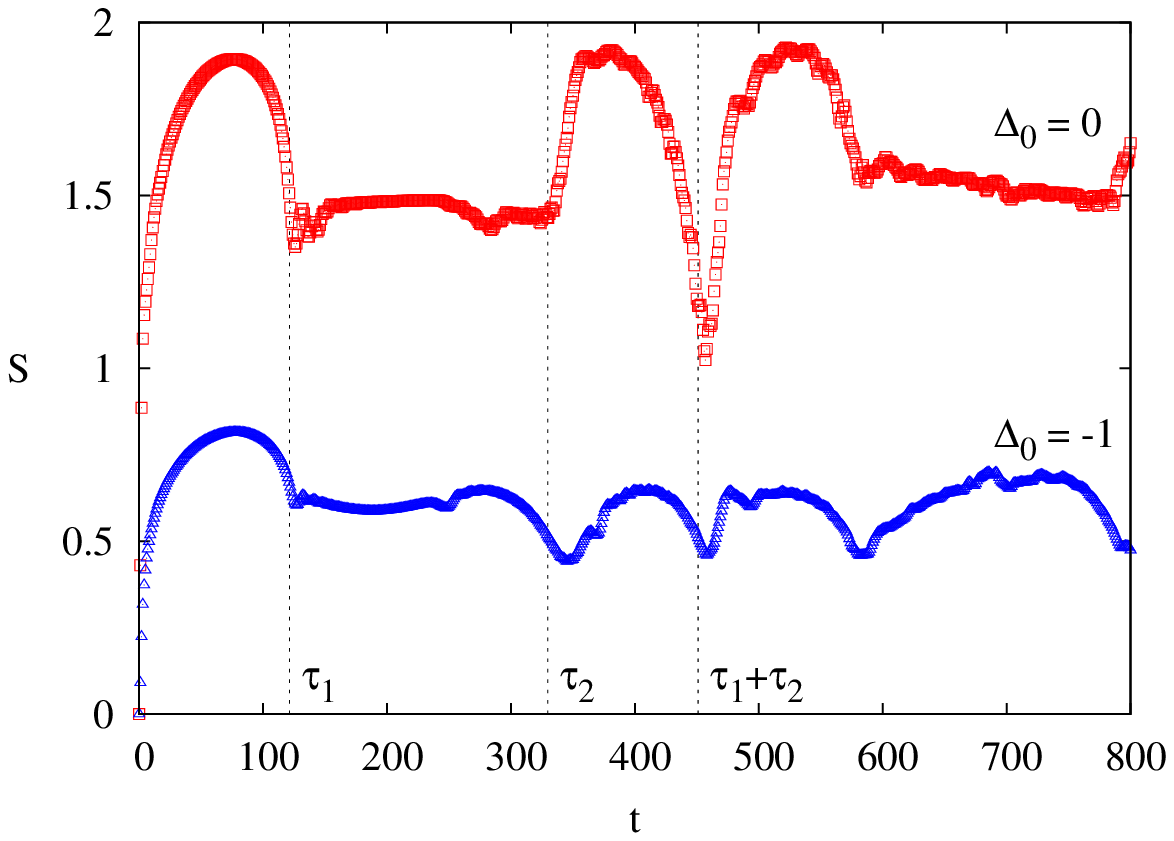}
\includegraphics[width=0.49\textwidth]{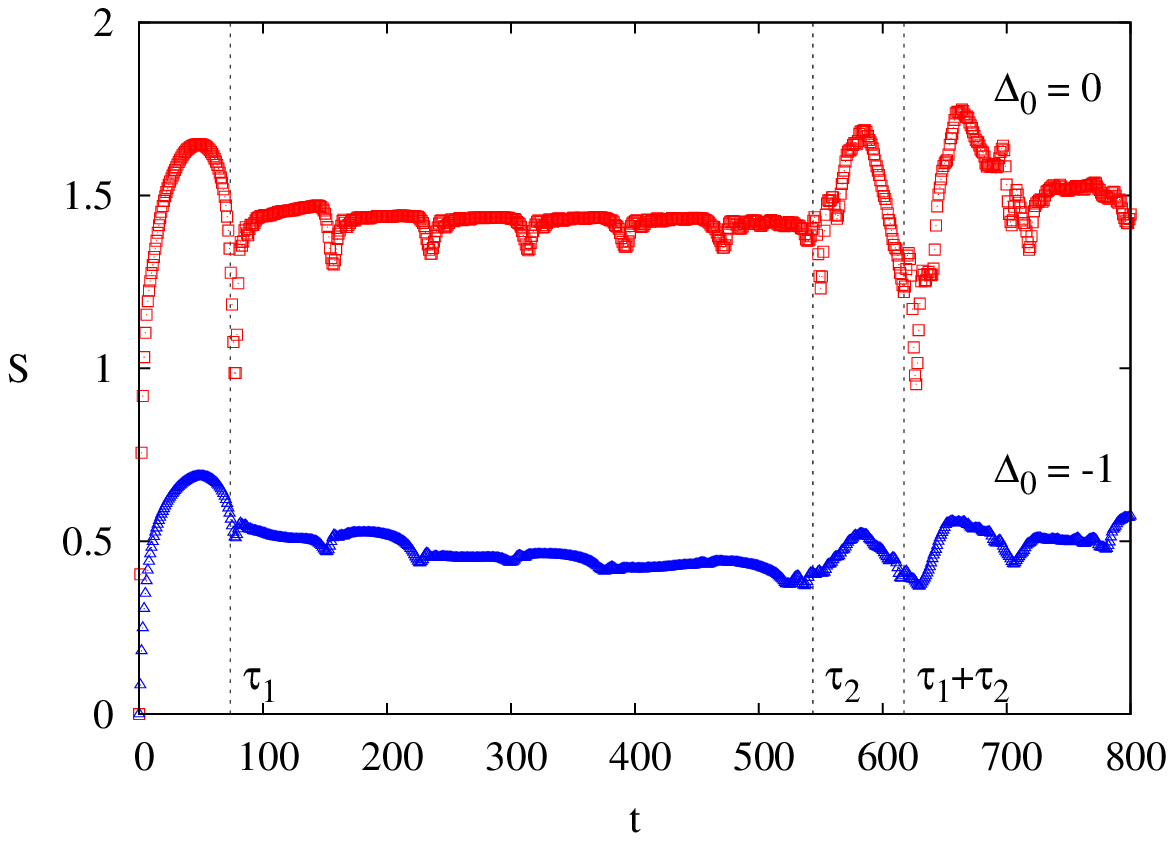}
\caption{Evolution of the entanglement entropy $S$ after connecting two metallic systems with
         $L=100$. Left: $\Delta=0.5$. Right: $\Delta=1$. The times indicated are the theoretical
         values, see the text.}
\label{ent_mm_quench}
\end{figure}

The two pieces are coupled at time $t=0$ by switching on $t_0$ and the evolution of
$C(t)$ is calculated via the Heisenberg operators $c_n(t),c^{\dag}_n(t)$.
The resulting $S(t)$ is shown in Fig. \ref{ent_mm_quench} for two different values of $\Delta$
and of $t_0$. The system on the left with $\Delta=0.5$ has roughly the band widths 
of Fig. \ref{fig:bands}, while on the right the asymmetry is larger.  

%
%
For $\Delta_0=0$ (upper curves) one has the following general features: An initial burst 
ending with a decrease at a time $\tau_1$ followed by a plateau, another burst at time
$\tau_2$ which is a kind of mirror image of the first ending at time $\tau_1+\tau_2$ and
then a rough repetition. For $\Delta=1$, the plateau is longer and shows six additional
structures, whereas for $\Delta=0.5$ only a single one is visible. The lower curves have
roughly the same pattern, but apart from the first maximum the structures are more washed 
out.

\begin{figure}[thb]
\centering
\includegraphics[scale=.4]{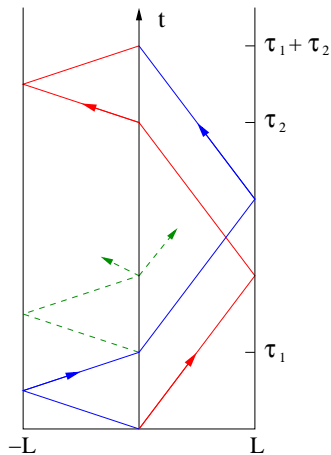}
\caption{Trajectories of the particles in the Calabrese-Cardy picture.}
\label{cc_picture}
\end{figure}

These features can be interpreted in the well-known picture due to Calabrese and Cardy
\cite{CC05}, in which a pair of particles is emitted from the junction at time $t=0$,
travelling in opposite directions and spreading the entanglement. In the present case, they 
have the velocities $v_1=t_1$ and $v_2=t_2$ and the corresponding space-time diagram is
given in Fig. \ref{cc_picture}.

%
%
After time $\tau_1=2L/v_1$, the left particle returns to the center, and if it simply moved
into the right subsystem, the entanglement would continue to drop. This happens in a 
somewhat related situation where the velocities are the same, but the size of the right
subsystem is larger \cite{Stephan/Dubail11}. Here, however, there is a probability for
reflection at the interface, such that the entanglement rises again and another small
``burst'' follows. For $\Delta=1$, where $v_1=7.4 \,v_2$, the left particle can make seven 
round trips before the right particle retuns to the center at time $\tau_2=2L/v_2=\tau_1/r$. These
are very clearly visible in the Figure. There are, however, some remarks to make.
According to (\ref{transmission_center_text}) there is no reflection right 
at the Fermi level for $\Delta_0=0$. The effect can therefore only come from somewhat 
slower particles away from it. In fact, the numerical values for $\tau_1$ and $\tau_2$
are somewhat larger than the theoretical ones, which supports this argument. 
Also, one would expect the reflection effect to become weaker with each
cycle, which is only barely the case. On the other hand, for $\Delta_0=-1$ one has $R=0.58$
at the Fermi level and the effect should be stronger, which is also not the case. Nevertheless, 
the picture gives a good overall description. 

One could presume that at least the first burst can be described by the conformal result
\cite{Stephan/Dubail11} 
\begin{equation}
S_h(vt)= \frac{c}{3}\ln{\left| \frac{2L}{\pi} \sin{\frac{\pi vt}{2L}} \right|}+ \mathrm{const.}
\label{conf_quench1} 
\end{equation}
valid for a homogeneous system with particle velocity $v$. Thus we tried the Ansatz
\begin{equation}
S(t)= [S_h(v_1t)+S_h(v_2t)]/2 + \mathrm{const.}
\label{conf_quench2} 
\end{equation}
which, on the scale of Fig. 12, fits the data for $\Delta_0=0$ very well. However, a closer look at 
the final decrease shows that it comes too early since, as mentioned, the value of $\tau_1$ is too 
small. Finally, we note that the first-burst results for a homogeneous but unequally divided system 
with the same round-trip times $\tau_1=2L_1$ and $\tau_2=2L_2$ lie above ours, and the difference 
increases with $\Delta$. Thus the two problems are not trivially connected.

\section{Summary}

We have considered the entanglement in fermionic chains composed of two different halves. 
In solid-state terminology, they were metals and insulators, while in statistical-physics 
terms they corresponded to critical and non-critical systems. The metal-metal system at
half filling showed the same logarithmic behaviour of $S$ as a homogeneous chain with a
defect. In the subleading terms, the asymmetry showed up, but the entropy had a remarkable 
scaling property. Its filling dependence, finally, showed oscillatory behaviour again linked 
to the asymmetry. Both features could be understood from the nature
of the extended eigenfunctions which are the same as in homogeneous systems with different
lengths.

The difference in the Fermi velocities showed up directly in the quench experiment where
two half-filled metals were joined and the entanglement was monitored. In that case, the value
of $r$ determined, how many cycles one sees in $S(t)$, before a certain return to the initial 
situation takes place. This pattern would occur in unsymmetrically divided homogeneous
systems only with additional defects.

The metal-insulator system was somewhat simpler. Its entanglement properties were seen to
lie always between the two pure systems and as a function of the size, one has a crossover
from an initial critical behaviour with a logarithmic increase of the entanglement to a
saturation typical for a non-critical system with finite correlation length. It also provided 
an instructive example for entanglement Hamiltonians which are quite different for the two
subsystems although they have the same spectra, as they must. One should note that the 
dimerized hopping model used for the insulator, and investigated already in \cite{Sirker14},
plays a central role in the theory of  polyacetylene \cite{Su/Schrieffer/Heeger79,Su/Schrieffer/Heeger80}.
In that sense, we studied a particular polymer system.   

We considered the geometry with open ends, because then one has only a single interface.
However, one could equally well look at rings. For composite transverse Ising models this was done
first in \cite{Hinrichsen90,Berche/Turban90} but without particular interface bonds and with a view on 
the spectra. Obtaining the entanglement entropy is more complicated, and we could only give
an analytical result for the asymptotic behaviour of the metal-metal system.

While we focussed on the entanglement entropy, it is known that the particle-number fluctuations 
in the subsystems behave similarly \cite{Song12}. Thus, in a homogeneous metal, the prefactor of 
the $\ln L$ term in the fluctuations is given by $T/2\pi^2$ with the transmission coefficient 
$T$ \cite{CMV12a} and varies qualitatively like $c_{\mathrm{eff}}$. The same result is found in the 
composite systems, and this could offer a way to access the entanglement experimentally.

\vspace{-0.2cm}

\begin{acknowledgements}
MCC and VE acknowledge the hospitality of Freie Universit\"at Berlin, where part of
this work was done. IP thanks National Tsing Hua University, Hsinchu, Taiwan for an invitation.
The work of VE was supported by OTKA Grant No.~NK100296 and MCC acknowledges NSC support under 
the contract No.~102-2112-M-005-001-MY3. 
\end{acknowledgements}

\section*{Appendix: Some formulae for the metal-metal system}

  Let $\bar\phi(n)$ and $\phi(n)$ denote the wave function at site $n$ on the 
left and right, respectively. The bulk solutions on the two sides are waves with momenta 
$q_1$ and $q_2$ and energy $\omega=-t_1\cos{q_1} = -t_2\cos{q_2}$. At the interface, the 
equations are
\begin{equation}
t_0\,\bar\phi(0) = t_2\,\phi(0) \quad, \quad t_1\,\bar\phi(1) = t_0\,\phi(1)
\label{interface} 
\end{equation}
For a wave coming in from the left and being partially reflected and transmitted, the
amplitudes are
\begin{equation}
\begin{split}
\bar\phi(n) &= A_1 \exp{(iq_1n)} + B_1 \exp{(-iq_1n)} \\
    \phi(n) &= A_2 \exp{(iq_2n)}
\label{scattering}
\end{split}
\end{equation}
Inserting this into the relations (\ref{interface}), one finds for the reflection
coefficient
\begin{equation}
R=\left|\frac{B_1}{A_1}\right|^2= \frac{\ch \, 2\nu-\cos{(q_1-q_2)}}{\ch \, 2\nu-\cos{(q_1+q_2)}}
\label{reflection} 
\end{equation}
where the quantity $\nu$ is defined as $\exp{(2\nu)}=t_0^2/t_1t_2$.
In our parametrization, where $t_1t_2=1$, one has $\nu=\Delta_0$. The transmission
coefficient follows from $T=1-R$ and can be written 
\begin{equation}
T = \frac{\sin{q_1}\sin{q_2}}{\sh^2 \, \nu+\sin^2{((q_1+q_2)/2)}}
\label{transmission} 
\end{equation}
If the left and right subsystems are identical, $q_1=q_2=q$, and one finds the result for a
bond defect in an otherwise homogeneous chain. The same holds in a general system if one is
in the middle of the bands. Then $\omega=0,q_1=q_2=\pi/2$ and one has the result
\begin{equation}
T = \frac{1}{\ch^2 \, \nu}
\label{transmission_center} 
\end{equation}
Due to the factor $\sin{q_2}$ in (\ref{transmission}), the transmission vanishes at the edges 
of the narrow band.

The extended eigenfunctions of $H$ have the form 
\begin{eqnarray}
\bar\phi(n) &=& A_1 \sin{q_1(n+L)} \nonumber \\
    \phi(n) &=& A_2 \sin{q_2(n-L-1)}
\label{eigenfunctions_h}
\end{eqnarray}
and the conditions (\ref{interface}) lead to
\begin{equation}
\frac{\sin{q_1(L+1)}\;\sin{q_2(L+1)}}{\sin{q_1L}\;\sin{q_2L}} = \frac{t_0^2}{t_1t_2}
\label{momenta1} 
\end{equation}
which, together with $t_1\cos{q_1}=t_2\cos{q_2}$, determines the allowed momenta.
Setting $q_{\alpha}=\pi/2+k_{\alpha}$ and assuming $L$ even, it takes the form 
\begin{equation}
\frac{\cos{k_1(L+1)}\;\cos{k_2(L+1)}}{\sin{k_1L}\;\sin{k_2L}} = \frac{t_0^2}{t_1t_2}
\label{momenta2} 
\end{equation}
The relation between the $k_{\alpha}$ reads
\begin{equation}
\sin{k_1}=r\sin{k_2}, \quad \quad r=t_2/t_1
\label{kmom} 
\end{equation}
and for small $r$ simplifies to $k_1=r\sin{k_2}$. Inserting this into (\ref{momenta2}) and 
setting $L+1 \simeq L$ in the numerator then gives
\begin{equation}
\cot{(k_2L)} = \frac{t_0^2}{t_1t_2} \tan{(rL\sin{k_2})}
\label{momenta3} 
\end{equation}
which contains only the momentum $k_2$.
The solutions can be discussed graphically, but one sees directly that besides $L$ a 
second length $L'=rL$ appears. If it is large enough, the tangent completes several 
cycles as $k_2$ varies, and this can lead to additional features in the momenta or in the 
amplitude ratio which is, again with $L+1 \simeq L$,
\begin{equation}
 \frac{A_2}{A_1} =  - \frac{t_1}{t_0}\; \frac{\cos(rL\sin{k_2})}{\sin{k_2L}} 
\label{amplitudes} 
\end{equation}
If $k_2$ is small, $k_1=rk_2$ for all $r<1$ and (\ref{momenta3}) for $t_0^2/t_1t_2 =1$ can be 
written in the two forms
\begin{equation}
\tan{(k_2rL)} \tan{(k_2L)} = 1, \quad \quad  \tan{(k_1L)} \tan{(k_1L/r)} = 1
\label{momenta4} 
\end{equation}
Both equations are the same as for a \emph{homogeneous} chain built up from two pieces 
with different lengths and can therefore be solved explicitly. For $k_2$, the homogeneous 
system has length $L+L'$, while for $k_1$ it has length $L+L''$, where $L''=L/r$. Thus
the small momenta are equidistant, i.e.
\begin{equation}
k_2 = \frac{\pi}{2(L+L')}\, (2n+1), \quad \quad n=0,\pm 1,\pm 2,...
\label{momenta5} 
\end{equation}
Using (\ref{momenta4}) in (\ref{amplitudes}), one finds $(A_2/A_1)^2=t_1^2=1/r$, i.e. the 
amplitudes are larger on the right than on the left. Including normalization to leading
order, this gives prefactors $ A_1^2 = 2/(L+L'')$ and $A_2^2 = 2/(L+L')$ which correspond 
exactly to these chain lengths.

\pagebreak

\section*{References}

\providecommand{\newblock}{}

\end{document}